\documentclass[conference,a4paper]{IEEEtran}
\IEEEoverridecommandlockouts
\IEEEpubid{\makebox[\columnwidth]{978-1-6654-7592-1/22/\$31.00~\copyright~2022 IEEE \hfill} \hspace{\columnsep}\makebox[\columnwidth]{ }}

\usepackage{cite}
\ifCLASSINFOpdf
\else
\fi

\usepackage{xcolor}
\usepackage{graphicx}
\usepackage{amsmath, bbm}
\usepackage{amssymb}
\usepackage{amsthm}
\usepackage{booktabs}
\usepackage{multirow}
\usepackage{array,multirow}
\usepackage{cite}
\usepackage{textcomp}
\usepackage{hyperref}
\usepackage{algorithm,algorithmic}
\usepackage{cancel}
\usepackage{tikz}
\usetikzlibrary{arrows, arrows.meta}
\usepackage{setspace}

\DeclareMathAlphabet\mathbfcal{OMS}{cmsy}{b}{n}

\begin{document}
%
\title{Frequency-aware Learned Image Compression \\ for Quality Scalability\vspace{-0.5em}}

\author{\IEEEauthorblockN{Hyomin Choi$^{\dag}$, Fabien Racapé$^{\dag}$, Shahab Hamidi-Rad$^{\dag}$, Mateen Ulhaq$^{\ast,\dag}$, Simon Feltman$^{\dag}$}
\IEEEauthorblockA{$^{\dag}$Interdigital Emerging Technologies Lab,
Los Altos, CA, USA\\
$^{\ast}$School of Engineering Science, Simon Fraser University, Burnaby, BC, Canada\\
Email: \{hyomin.choi; fabien.racape; shahab.hamidi-rad; mateen.ulhaq; simon.feltman\}@interdigital.com}}

\maketitle
\IEEEpubidadjcol

\begin{abstract}
Spatial frequency analysis and transforms serve a central role in most engineered image and video lossy codecs, but are rarely employed in neural network (NN)-based approaches. 
We propose a novel NN-based image coding framework that utilizes forward wavelet transforms to decompose the input signal by spatial frequency.
Our encoder generates separate bitstreams for each latent representation of low and high frequencies. This enables our decoder to selectively decode bitstreams in a quality-scalable manner.
Hence, the decoder can produce an enhanced image by using an enhancement bitstream in addition to the base bitstream. Furthermore, our method is able to enhance only a specific region of interest (ROI) by using a corresponding part of the enhancement latent representation. Our experiments demonstrate that the proposed method shows competitive rate-distortion performance compared to several non-scalable image codecs. We also showcase the effectiveness of our two-level quality scalability, as well as its practicality in ROI quality enhancement. 
\end{abstract}


\renewcommand\IEEEkeywordsname{Index Terms}

%
\IEEEpeerreviewmaketitle

\section{Introduction}

Conventional lossy codecs~\cite{sayood2017introduction} use 
transforms such as the Discrete Wavelet Transform (DWT) or Discrete Cosine Transform (DCT), alongside quantization to achieve variable-rate compression.
An image is transformed into the frequency domain,
and the resulting transformed coefficients are quantized into bins, where each bin is sized to minimize perceptible distortion. Most importantly, distortion at high spatial frequencies is much less noticeable than distortion at low frequencies.
Based on this property of the human visual system,  
a quality-scalable coding method allows progressive improvement in decoded image quality by providing the decoder with further high-frequency information at higher bitrates~\cite{shapiro1993embedded, christopoulos2000jpeg2000, schwarz2007overview, boyce2015overview}.

End-to-end learned image compression (LIC) methods have recently caught the research community's interest.
Ball{\'e} \textit{et al.}~\cite{balle2015density} first proposed a density modeling method using Generalized Divisive Normalization (GDN) 
to \emph{transform} the input images into an entropy coding-friendly latent space, which was effectively used in the autoencoder-based~\cite{Goodfellow-et-al-2016} approach in \cite{balle2016end}.
More recently, several variational autoencoder (VAE)-based methods~\cite{balle2018variational, minnen2018joint, cheng2020image} focused on accurately modeling the distributions of the latent variables, resulting in rate-distortion (RD) performance competitive with the latest fully-engineered codecs~\cite{hevc_std, vvc_std}.
Other approaches~\cite{xie2021enhanced, akbari2021learned} sought improvements in the analysis transform%
\footnote{In LIC literature, the analysis transform $g_a$ transforms an input image into the latent space, from which the synthesis transform $g_s$ reconstructs the image.}.
In particular, Akbari \textit{et al.}~\cite{akbari2021learned} replaced regular 2D-convolutions with octave convolutions (OctConv)~\cite{chen2019drop} which act like wavelet transforms in that the spatial resolution is reduced while diminishing spatial redundancy.
However, rather than analyzing the input by frequency, the authors' modifications to OctConv focus on improving representational power. 

We propose a quality-scalable frequency-aware learned image coding (FLIC) method using wavelet-embedded octave convolution (WeOctConv)
that has RD performance competitive with non-scalable methods.
Our method supports two-level quality scalability by encoding an input image into two separate bitstreams, as well as ROI-based quality scalability by encoding only selected regions of the latent space.

In Section~\ref{sec:prior_work}, we review prior work that inspired the proposed method, which is then described in detail in Section~\ref{sec:propose_method}. Experimental results are presented in Section~\ref{sec:experiments}, followed by conclusions in Section~\ref{sec:conclusion}.

\section{Prior work}
\label{sec:prior_work}

Chen~\textit{et al.}~\cite{chen2019drop} introduced the OctConv layer which factorizes its input into low-frequency ($L$) and high-frequency ($H$) features.
Given an input tensor $\mathbfcal{Y}_{\textup{in}}=\{\mathbfcal{Y}_{\textup{in}}^{H}, \mathbfcal{Y}_{\textup{in}}^{L}\}$, the output tensor $\mathbfcal{Y}_{\textup{out}}=\{\mathbfcal{Y}_{\textup{out}}^{H}, \mathbfcal{Y}_{\textup{out}}^{L}\}$ is computed by
\begin{equation}
\begin{split}
  & \mathbfcal{Y}_{\textup{out}}^{H}
  = f^{H \rightarrow H}(
    \mathbfcal{Y}_{\textup{in}}^{H}
  )
  + \texttt{Upsample}(
    f^{L \rightarrow H}(
      \mathbfcal{Y}_{\textup{in}}^{L}
    )
  ) \\
  & \mathbfcal{Y}_{\textup{out}}^{L}
  = f^{L \rightarrow L}(
    \mathbfcal{Y}_{\textup{in}}^{L}
  )
  + f^{H \rightarrow L}(
    \texttt{Pool}(
      \mathbfcal{Y}_{\textup{in}}^{H}
    )
  )
\end{split}
\end{equation}
\noindent
where $f^{A\rightarrow B}$ represents the convolutional update between groups of frequencies $A$ and $B$.
$f^{A\rightarrow B}$ is known as \emph{inter-frequency communication} whenever $A$ and $B$ are distinct frequency groups, and \emph{intra-frequency update} whenever they are the same. 
\texttt{Upsample} uses nearest-neighbor interpolation.
\texttt{Pool} uses average pooling, which from the perspective of frequency analysis is similar to a low-pass filter for the 2-D discrete Haar wavelet transform.
Thus, $\mathbfcal{Y}_{\textup{out}}^{L}$ accumulates the low-frequency information over successive OctConv layers.

Akbari~\textit{et al.}~\cite{akbari2021learned} introduced a modified OctConv in which \texttt{Pool} is replaced with a convolution with a stride of 2.
This modification increases representational ability, but is also less interpretable from the frequency decomposition perspective.
Additionally, extensively applying GDN to the OctConv layers potentially adds redundant computations since OctConv already conducts some degree of factorization.
Nonetheless, their approach shows improvements in RD performance in comparison to several LIC approaches.


\section{Proposed method}
\label{sec:propose_method}
This section describes the proposed WeOctConv layer and introduces our quality-scalable FLIC method, consisting of WeOctConvs, based on the factorized prior architecture~\cite{balle2016end}.


\begin{figure}[t!]
    \centering
    \begin{minipage}[b]{0.48\linewidth}
    \centering
    \includegraphics[width=\textwidth]{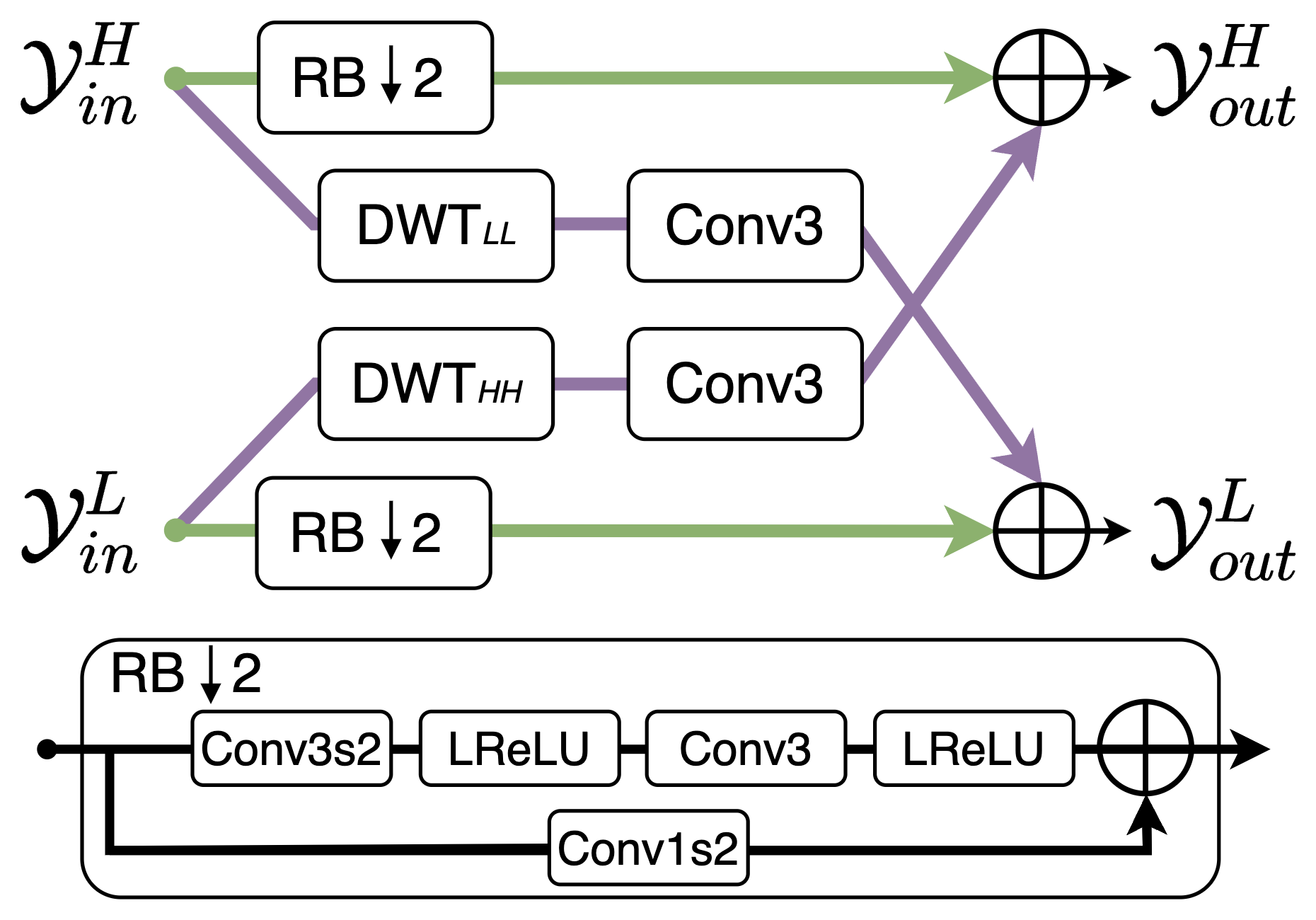}
    \centerline{(a)}\medskip
    \end{minipage}
    \hspace{0.1cm}
    \centering
    \begin{minipage}[b]{0.48\linewidth}
    \centering
    \includegraphics[width=\textwidth]{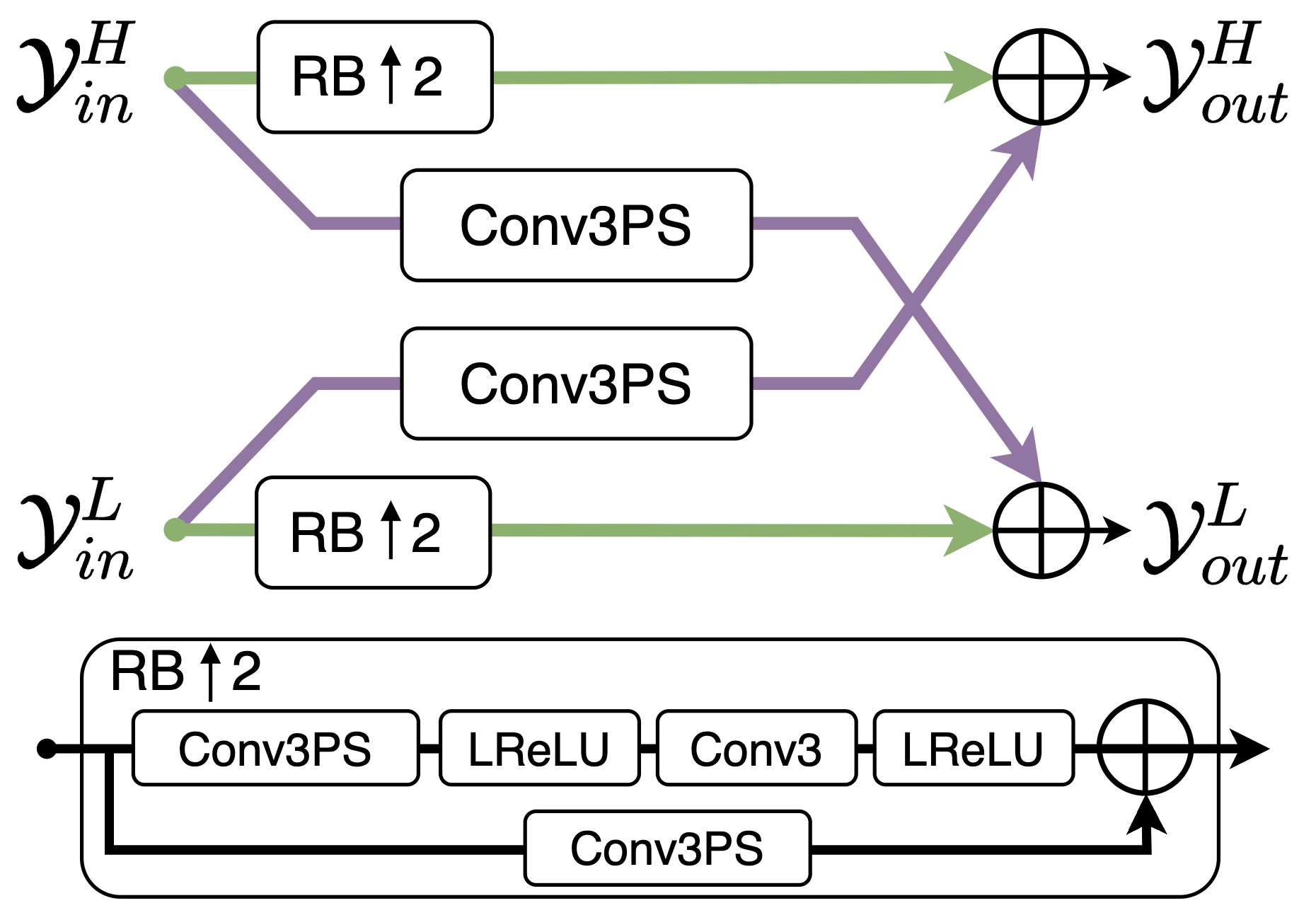}
    \centerline{(b)}\medskip
    \end{minipage}
\vspace{-0.6cm}
\caption{%
  Design of the proposed layers (a) WeOctConv and (b) TWeOctConv, along with their corresponding Residual Block (RB).
  Inter-frequency updates are shown as purple lines, and intra-frequency updates as green lines.%
}
\vspace{-0.35cm}
\label{fig:weoctconv}
\end{figure}

\subsection{Wavelet-embedded Octave Convolution}
Fig.~\ref{fig:weoctconv} depicts the overall computation flow of the proposed WeOctConv layer, used in the analysis transform, and its dual TWeOctConv, used in the synthesis transform.
The WeOctConv layer shown in Fig.~\ref{fig:weoctconv}(a) captures low and high frequency information from input tensors $\mathbfcal{Y}_{in}^{k} \in \mathbb{R}^{C_{in} \times N \times M}$ for each $k\in\{L, H\}$, where $C_{in}$ is the number of input channels and $N \times M$ is the feature resolution, and outputs the tensors $\mathbfcal{Y}_{out}^{k} \in \mathbb{R}^{C_{out} \times \frac{N}{2} \times \frac{M}{2}}$, where $C_{out}$ is the number of output channels.
Conversely, the TWeOctConv layer in Fig.~\ref{fig:weoctconv}(b) synthesizes $\mathbfcal{Y}_{in}^{k} \in \mathbb{R}^{C_{in} \times \frac{N}{2} \times \frac{M}{2}}$ into $\mathbfcal{Y}_{out}^{k} \in \mathbb{R}^{C_{out} \times N \times M}$.

For the WeOctConv layer, the inter-frequency update consists of a DWT and a convolution, where
the DWTs use the following $2\times 2$ Haar wavelet kernels with a stride of 2:
\begin{equation}
LL= \frac{1}{2}
\begin{bmatrix}
1 & 1\\ 
1 & 1
\end{bmatrix}, \quad HH=\frac{1}{2}
\begin{bmatrix}
\phantom{-}1 & -1\\ 
-1 &  \phantom{-}1
\end{bmatrix}.
\label{eq:haar_dwt}
\end{equation}
\noindent
The low-pass filter $LL$ is used for the $H\rightarrow L$ update, and the high-pass filter $HH$ is used for the $L\rightarrow H$ update.
The downsampled filtered output is then fed into a convolutional layer with $3\times3$ kernels, denoted as Conv3.
We found that it is challenging to simultaneously optimize for both the trainable DWT transform coefficients~\cite{wolter2021adaptive} and the RD criterion.
Hence, we instead use fixed DWT transform coefficients.
For the TWeOctConv layer, the natural choice for the inter-frequency update is a Conv3 and an inverse DWT (IDWT).
However, we found that using fixed IDWT transform coefficients causes severe quality degradation, especially at lower bitrates.
Hence, we instead use a Conv3PS layer composed of a Conv3 followed by a PixelShuffle~\cite{shi2016real}, which is capable of upsampling its input in a way similar to IDWT by mixing groups of 4 channels.

Each intra-frequency update uses Residual Blocks (RBs).
For WeOctConv, each RB consists of a Conv3 with stride 2 (denoted as Conv3s2), a L(eaky)ReLU, a Conv3, and a LReLU; and in the ``skip'' branch, a convolution with a $1\times1$ kernel and stride 2 (denoted as Conv1s2).
For TWeOctConv, Conv3PS replaces both Conv3s2 and Conv1s2 within each RB.

Lastly, the inter-frequency updates are added to their corresponding intra-frequency updates. 

\begin{figure}[t]
    \centering
    \begin{minipage}[b]{1\linewidth}
    \centering
    \includegraphics[width=\textwidth]{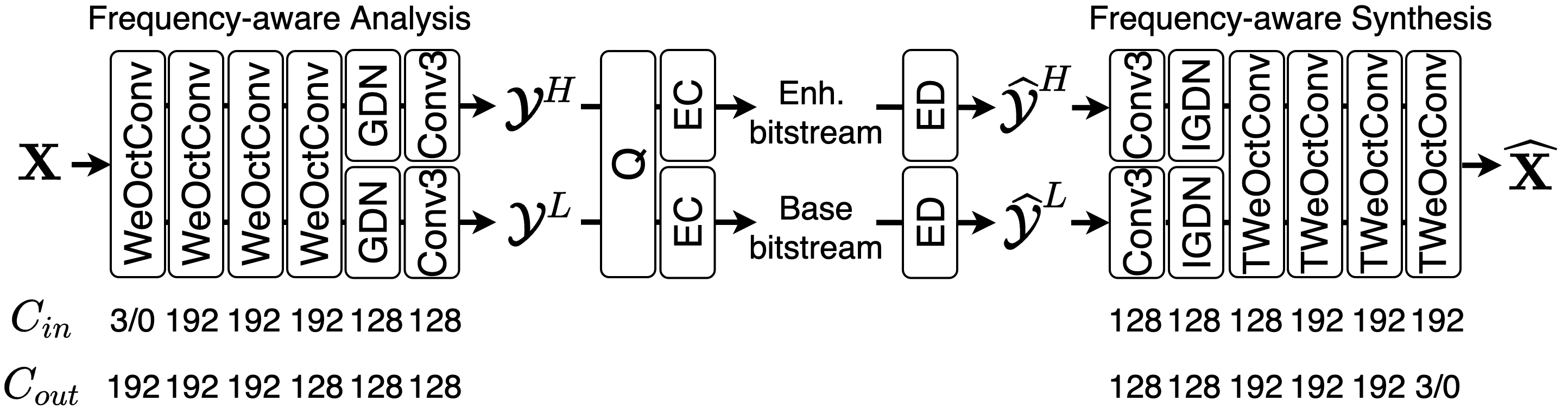}
    \end{minipage}
    \vspace{-.3cm}
\caption{Overall architecture of the proposed FLIC. The bottom of each layer details a number of input channels $C_{in}$ and output channels $C_{out}$, respectively. For the first layer $C_{in}$ in the analysis and the last layer $C_{out}$ in the synthesis, only one port of input and output with three channels (i.e., RGB) is available. }
\label{fig:flic_architecture}
\end{figure}

\subsection{Frequency-aware Learned Image Compression}
Fig.~\ref{fig:flic_architecture} presents the overall architecture of the proposed image compression network employing (T)WeOctConvs. 
Denoted by ``Frequency-aware Analysis'', the encoder-side analysis transform $g_a(\cdot)$ with learned parameters $\psi_{a}$ analyzes an input image $\mathbf{X}$ and generates a pair of compressed latent representations:
\begin{equation}
    \mathbfcal{Y} = \{\mathbfcal{Y}^{L}, \mathbfcal{Y}^{H}\} = g_{a}(\mathbf{X};\psi_{a}).
\end{equation}
We limit the use of GDNs to only one instance after the last WeOctConv. 
In our experiments, reducing the number of computationally intensive GDNs in this way does not result in a performance drop.
Furthermore, no bias is used at all to reduce the number of operations.
$\mathbfcal{Y}^{H}$ and $\mathbfcal{Y}^{L}$ are then rounded to the nearest integer during inference to obtain the quantized latent representations $\widehat{\mathbfcal{Y}}^{L}$ and $\widehat{\mathbfcal{Y}}^{H}$.
During training, like in~\cite{balle2016end}, uniform noise is added to the latent representations for the gradient computation.
$\widehat{\mathbfcal{Y}}^{L}$ and $\widehat{\mathbfcal{Y}}^{H}$ are respectively losslessly encoded into a \emph{base} and an \emph{enhancement} bitstream using entropy encoders (ECs)%
\footnote{Specifically, arithmetic range coder based on Asymmetric Numeral Systems (ANS) provided in~\cite{begaint2020compressai} is used.}.

The decoder uses entropy decoders (EDs) to losslessly decode the given bitstreams into $\widehat{\mathbfcal{Y}}^{L}$ and $\widehat{\mathbfcal{Y}}^{H}$.
Denoted by ``Frequency-aware Synthesis'', the synthesis transform $g_s(\cdot)$ with learned parameters $\psi_s$ synthesizes a reconstructed image $\widehat{X}$ from $\widehat{\mathbfcal{Y}}^{L}$ and $\widehat{\mathbfcal{Y}}^{H}$ by applying to each a Conv3 and an inverse GDN (IGDN), and then applying a number of TWeOctConvs.
This is written as 
\begin{equation}
    \widehat{\mathbf{X}} = g_s(\widehat{\mathbfcal{Y}}^{L}, \widehat{\mathbfcal{Y}}^{H};\psi_s).
\label{eq:full_quality}
\end{equation}

\begin{figure}[t]
    \centering
    \begin{minipage}[b]{.30\linewidth}
    \centering
    \includegraphics[width=\textwidth]{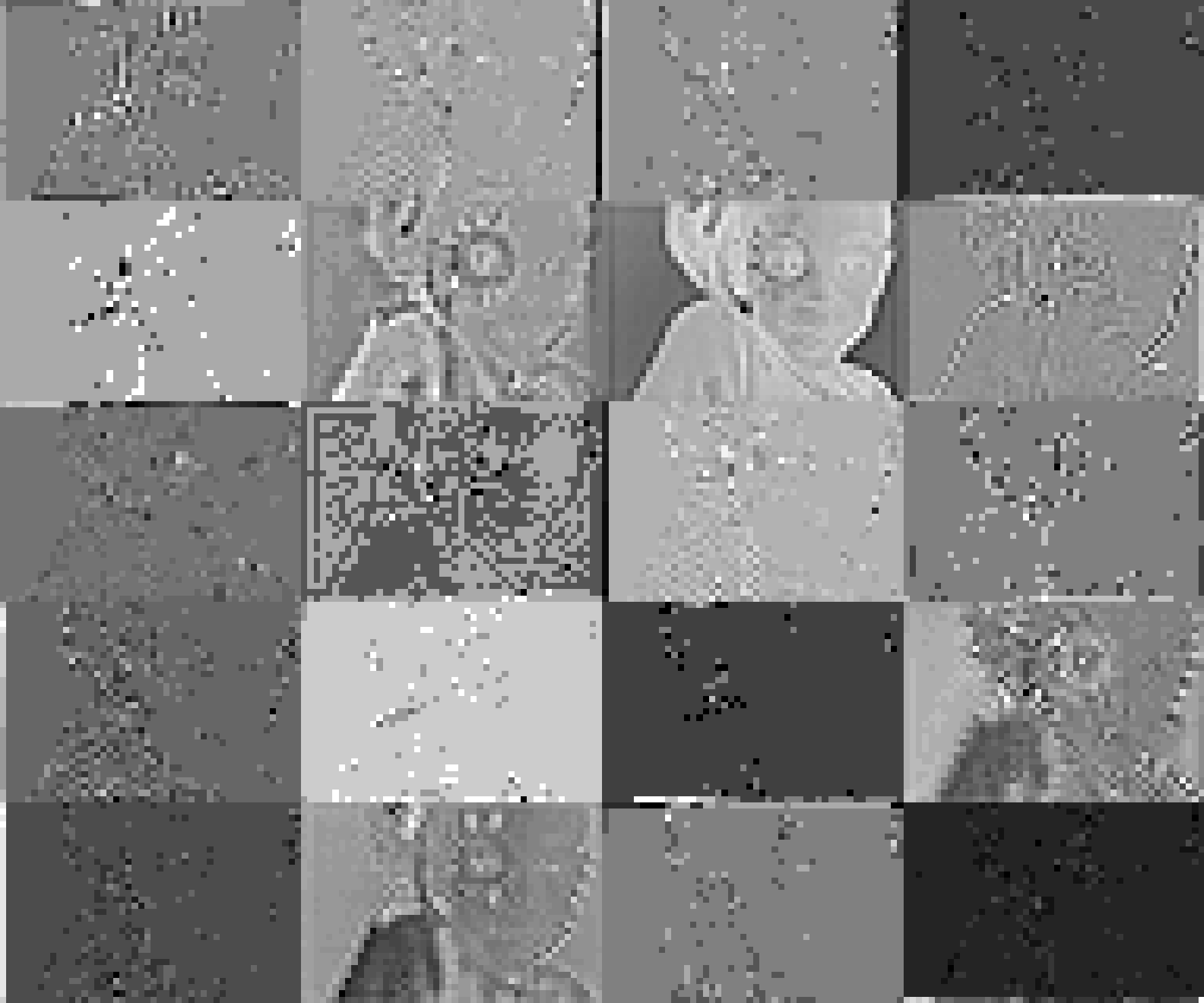}
    \centerline{(a)}\medskip
    \end{minipage}
    \centering
    \hspace{0.02cm}
    \begin{minipage}[b]{.67\linewidth}
    \centering
    \includegraphics[width=\textwidth]{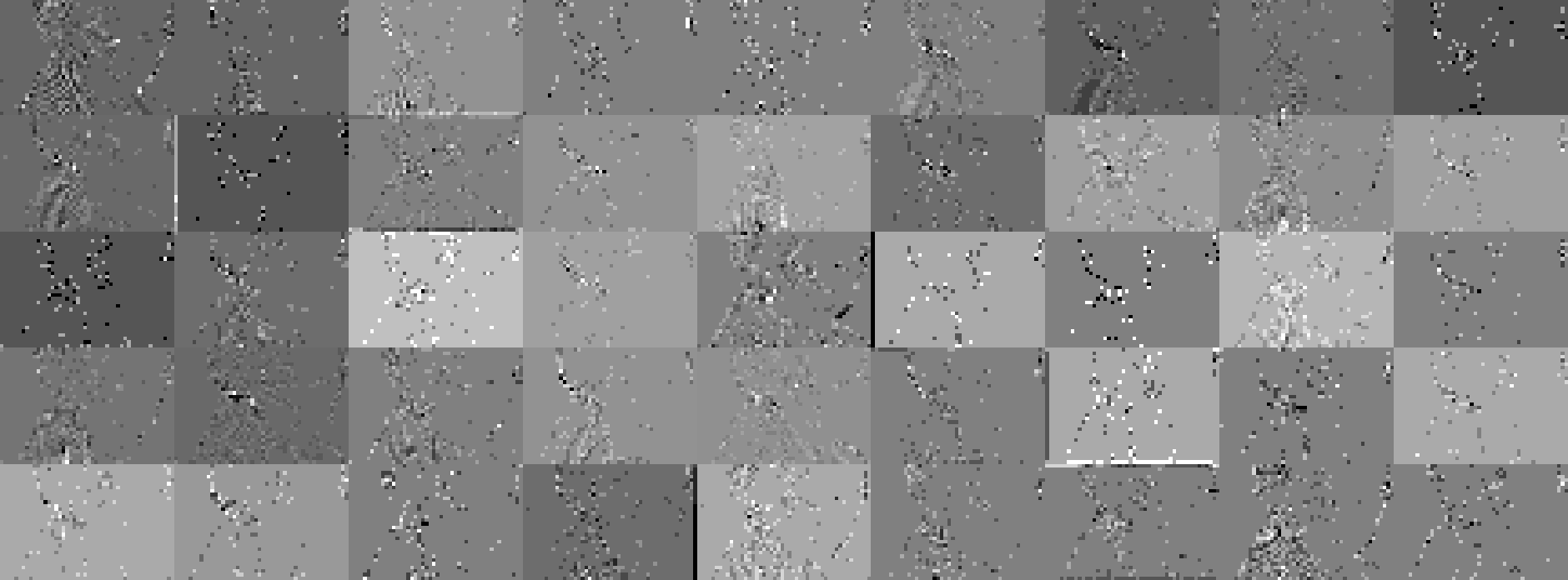}
    \centerline{(b)}\medskip
    \end{minipage}
\vspace{-0.6cm}
\caption{Tiled latent channels of (a) $\widehat{\mathbfcal{Y}}^{L}$ and (b) $\widehat{\mathbfcal{Y}}^{H}$ computed using a sample image from the Kodak dataset~\cite{kodak_dataset}. Each channel is normalised to $[0,1]$ for visibility. All channels with non-zero latent representations are presented.}
\vspace{-0.3cm}
\label{fig:tiled_tensor}
\end{figure}

As shown in Fig.~\ref{fig:tiled_tensor}, spatial low- and high-frequency features are captured well by $\widehat{\mathbfcal{Y}}^{L}$ and $\widehat{\mathbfcal{Y}}^{H}$, respectively.
This is because WeOctConv examines the wavelet decomposition on each layer's input throughout the analysis transform.

\subsection{Quality scalability}
In traditional scalable codecs~\cite{shapiro1993embedded, christopoulos2000jpeg2000, schwarz2007overview}, the decoder may reconstruct an input image using its encoded base bitstream.
A more detailed, higher quality image may be reconstructed by additionally providing the decoder with an enhancement bitstream that typically carries further high-frequency information.
The proposed FLIC provides a similar form of quality scalability.
Eq.~(\ref{eq:full_quality}) generates a high-quality input reconstruction using all latent representations.
However, a lower quality image may be reconstructed by setting $\widehat{\mathbfcal{Y}}^{H} = \mathbf{0}$ so that
\begin{equation}
    \widehat{\mathbf{X}}_{\textup{base}} = g_s(\widehat{\mathbfcal{Y}}^{L}, \mathbf{0};\psi_s),
\label{eq:base_quality}
\end{equation}
\noindent where $\mathbf{0}$ denotes a zero tensor whose elements are all zeros with the same dimension as $\widehat{\mathbfcal{Y}}^{H}$.

Furthermore, the decoder also supports the quality enhancement of selected ROIs.
This can be done by feeding $g_s$ with a decoded tensor $\widehat{\mathbfcal{Y}}^{H}_{\textup{ROI}}$ (of the same dimensions as $\widehat{\mathbfcal{Y}}^{H}$) containing coded latent variables only for corresponding regions and zeros everywhere else.
This produces the reconstructed image
\begin{equation}
    \widehat{\mathbf{X}}_{\textup{ROI}} = g_s(\widehat{\mathbfcal{Y}}^{L}, \widehat{\mathbfcal{Y}}^{H}_{\textup{ROI}};\psi_s),
\label{eq:roi_quality}
\end{equation}
\noindent  
where the selected ROI regions have been enhanced.

\subsection{Loss function}
During training, we use a loss function in the form of an RD Lagrangian as in~\cite{balle2016end}:
\begin{equation}
\mathcal{L} = \underbrace{\mathbb{E}_{x\sim p_x}\left [ -\textup{log}_{2}p_{\hat{y}}(\hat{y}) \right ]}_{\textup{rate estimation}} + \lambda \cdot (\underbrace{D(\mathbf{X},\widehat{\mathbf{X}}) + \alpha \cdot D(\mathbf{X},\widehat{\mathbf{X}}_{\textup{base}})}_{\textup{distortion}}),
\label{eq:loss}
\end{equation}
\noindent where $p_x$ denotes the probability density of the input data $x$ and $p_{\hat{y}}$ represents a fully factorized distribution of the quantized latent variable $\hat{y}$.
The distortion metric $D$ can be any objective quality metric;
we use mean squared error (MSE) and MS-SSIM~\cite{wang2003multiscale} for our experiments.
The hyperparameter $\alpha \ge 0$ balances the importance in quality of the full base+enhancement reconstruction $\widehat{\mathbf{X}}$ and the base-only reconstruction $\widehat{\mathbf{X}}_{\textup{base}}$.

\section{Experiments}
\label{sec:experiments}

Our FLIC networks are trained on random cropped patches of size $256 \times 256$ from the Vimeo-90K dataset~\cite{xue2019video}.
The mini-batch size is set to 8 and our networks are trained for up to 2.5M steps ($\approx$ 350 epochs), corresponding to about 10 to 12 days.
We use an Adam optimizer with an initial learning rate of $10^{-4}$, which is then decreased by 90\% after the first 30 epochs whenever the validation loss plateaus with a patience of 4 epochs.
We train models for each $\lambda=2^n \cdot 10^{-2}$ over all $n \in \{3, 2, 1, 0, -1, -2, -3\}$. 
The models are tested on all 24 images in the Kodak dataset~\cite{kodak_dataset} to evaluate RD performance.


\subsection{Compression performance}

\begin{figure}[t]
    \centering
    \begin{minipage}[b]{0.86\linewidth}
    \centering
    \includegraphics[width=\textwidth]{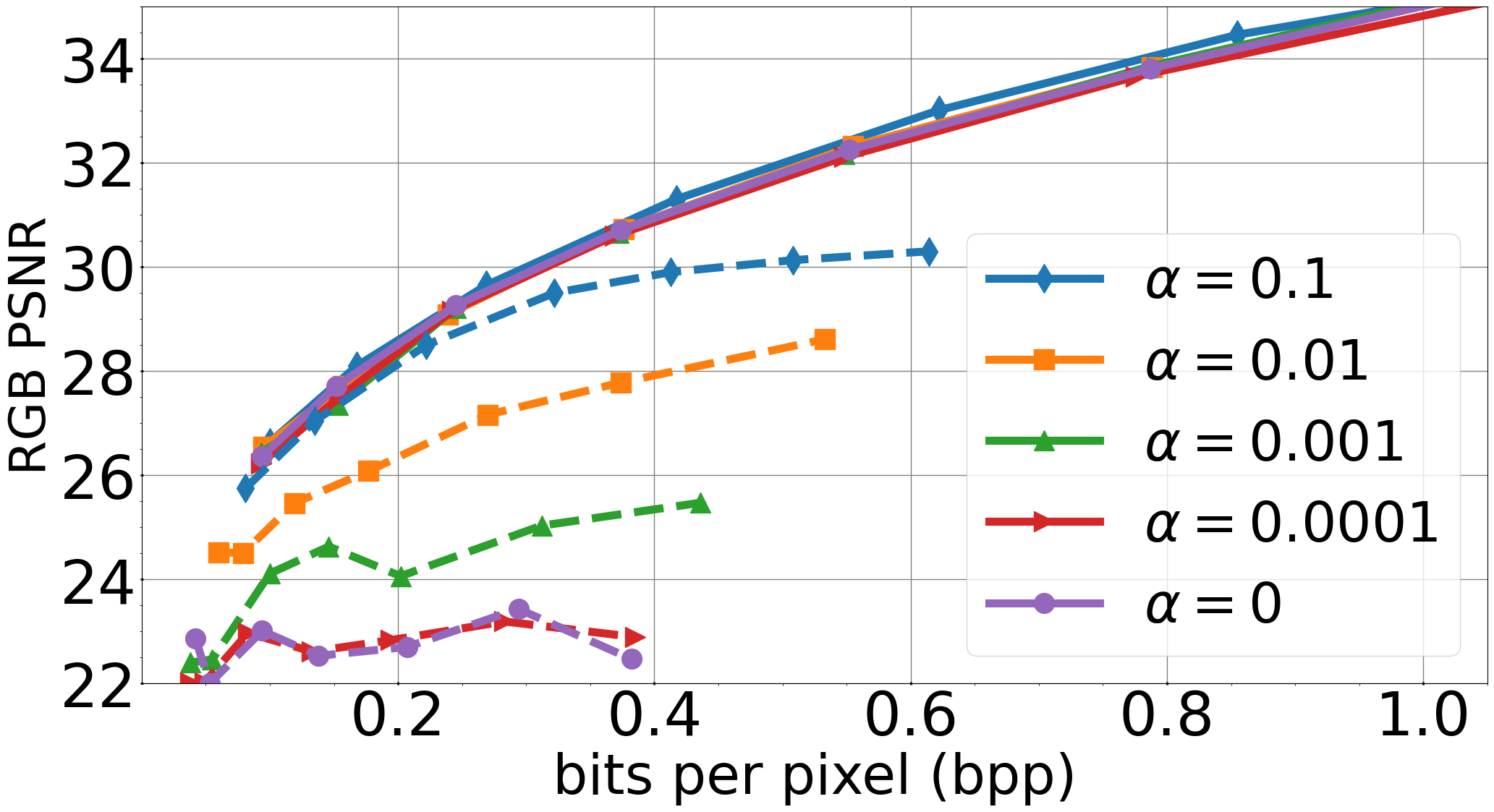}
    \end{minipage}
\vspace{-0.2cm}    
\caption{RD performance of our proposed FLIC with various $\alpha$ on the Kodak dataset~\cite{kodak_dataset}. Solid and dashed lines represent the full bitstream and the base-only bitstream, respectively.}
\vspace{-.2cm}
\label{fig:alpha_impact}
\end{figure}

\begin{figure}[t]
    \centering
    \begin{minipage}[b]{0.95\linewidth}
    \centering
    \includegraphics[width=\textwidth]{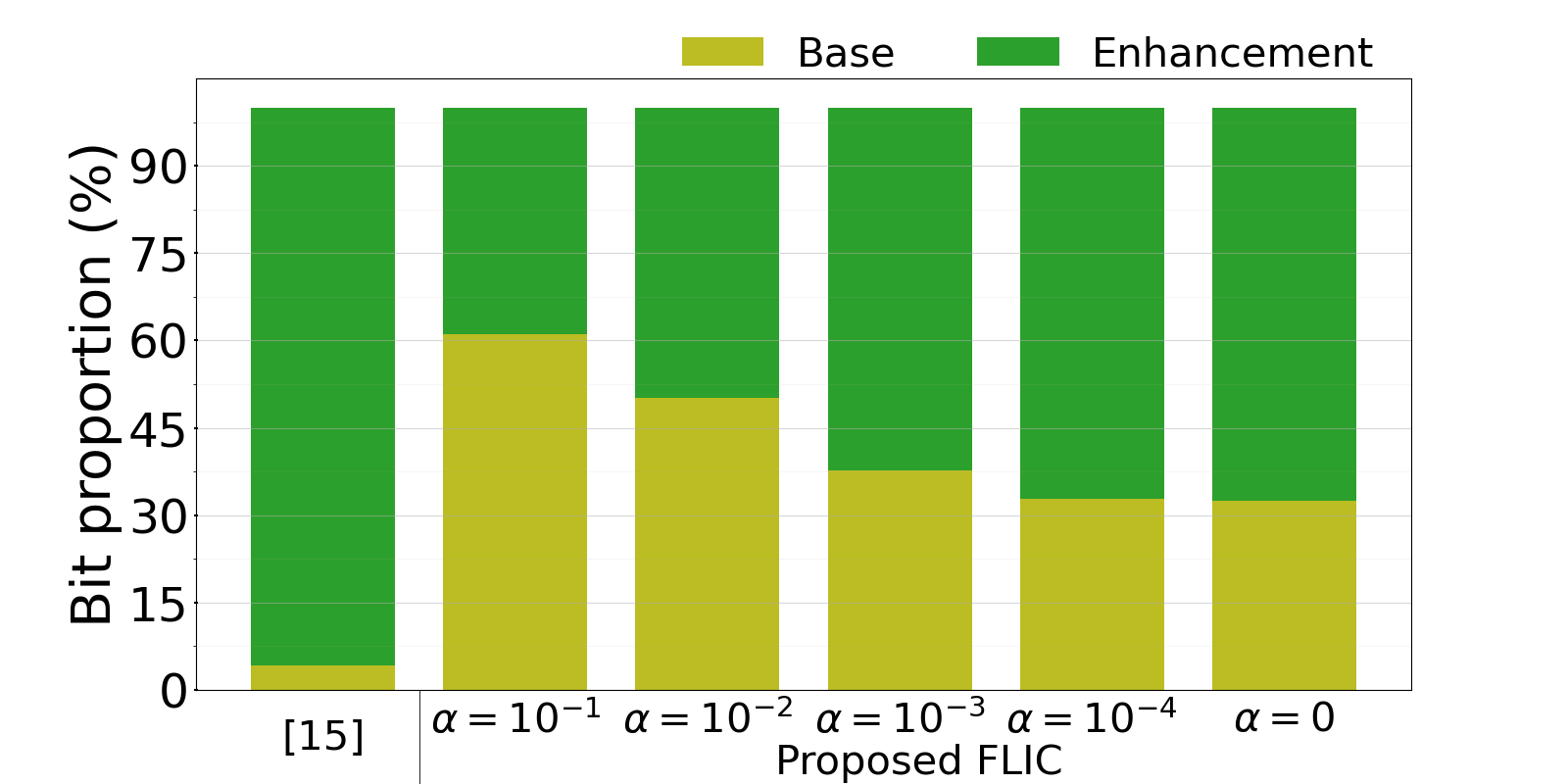}
    \end{minipage}
\vspace{-0.2cm}
\caption{Comparison of average bit proportion between the base layer and the enhancement layer for several models including~\cite{akbari2021learned} and the proposed models, trained with $\lambda=2^{3} \cdot 10^{-2}$.}
\vspace{-0.3cm}
\label{fig:bpp_proportion}
\end{figure}

\subsubsection{Impact of the hyperparameter \texorpdfstring{$\alpha$}{alpha}} 
We introduced $\alpha$ in Eq.~(\ref{eq:base_quality}) to control the quality of $\widehat{\mathbf{X}}_{\textup{base}}$.
To examine its impact on RD performance, we trained our model over various values of $\lambda$ for several $\alpha \in \{0.1, 0.01, 0.001, 0.0001, 0\}$.
The RD curves for various $\alpha$ are shown in Fig.~\ref{fig:alpha_impact}.
The RD performance of the full bitstream (solid curves) is only marginally affected by changing $\alpha$. 
In contrast, the RD performance of the base-only bitstream (dashed curves) drastically improves as $\alpha$ increases.
Thus, a larger value of $\alpha \ge 0.1$ may be used to obtain good base-only reconstructions without compromising full reconstruction performance.

Fig.~\ref{fig:bpp_proportion} compares the average bit proportion between the base and enhancement bitstreams for various models. 
Although there is no scalability in~\cite{akbari2021learned}%
\footnote{Since there is no publicly available code, we have reimplemented the network using the factorized prior-based entropy bottleneck~\cite{balle2016end} and trained it on the same dataset used for our proposed model.},
we consider their low- and high-frequency feature tensors to be the base and the enhancement layers, respectively.
On average, about 4\% of the entire bitstream by~\cite{akbari2021learned} represents low-frequency information.
In contrast, our method adaptively uses $\alpha$ to control the bit proportion trade-off between bitstreams, without significantly affecting overall RD performance.

\begin{table}[t]
\centering
\renewcommand{\arraystretch}{1.25}  
\caption{BD-rate (\%) performance of various learned image codecs against JPEG2000~\cite{christopoulos2000jpeg2000}}
\label{tbl:bd_performance}
\smallskip\noindent
\resizebox{1.0\linewidth}{!}{%
\begin{tabular}{c|cc|ccc}
\hline
Anchor     & \multicolumn{2}{c|}{Benchmark LIC models}      & \multicolumn{3}{c}{Proposed models}       \\ 
JPEG2000   & \cite{balle2016end} & \cite{akbari2021learned} & $\alpha=0.1$ & $\alpha=0.01$ & $\alpha=0$ \\ \hline
opt-MSE    & -17.48              & -14.48                   & -16.11       & -12.72        & -13.24     \\ \hline
opt-MS-SSIM & -62.48              & -64.88                   & -63.30       & -62.51        & -63.73     \\ \hline 
\end{tabular}}
\vspace{-0.2cm}
\end{table}

\begin{figure*}[!htb]
    \centering
    \begin{minipage}[b]{0.21\linewidth}
    \centering
    \centerline{\small Original (uncompressed)}
    \vspace{0.1cm}
    \includegraphics[width=\textwidth]{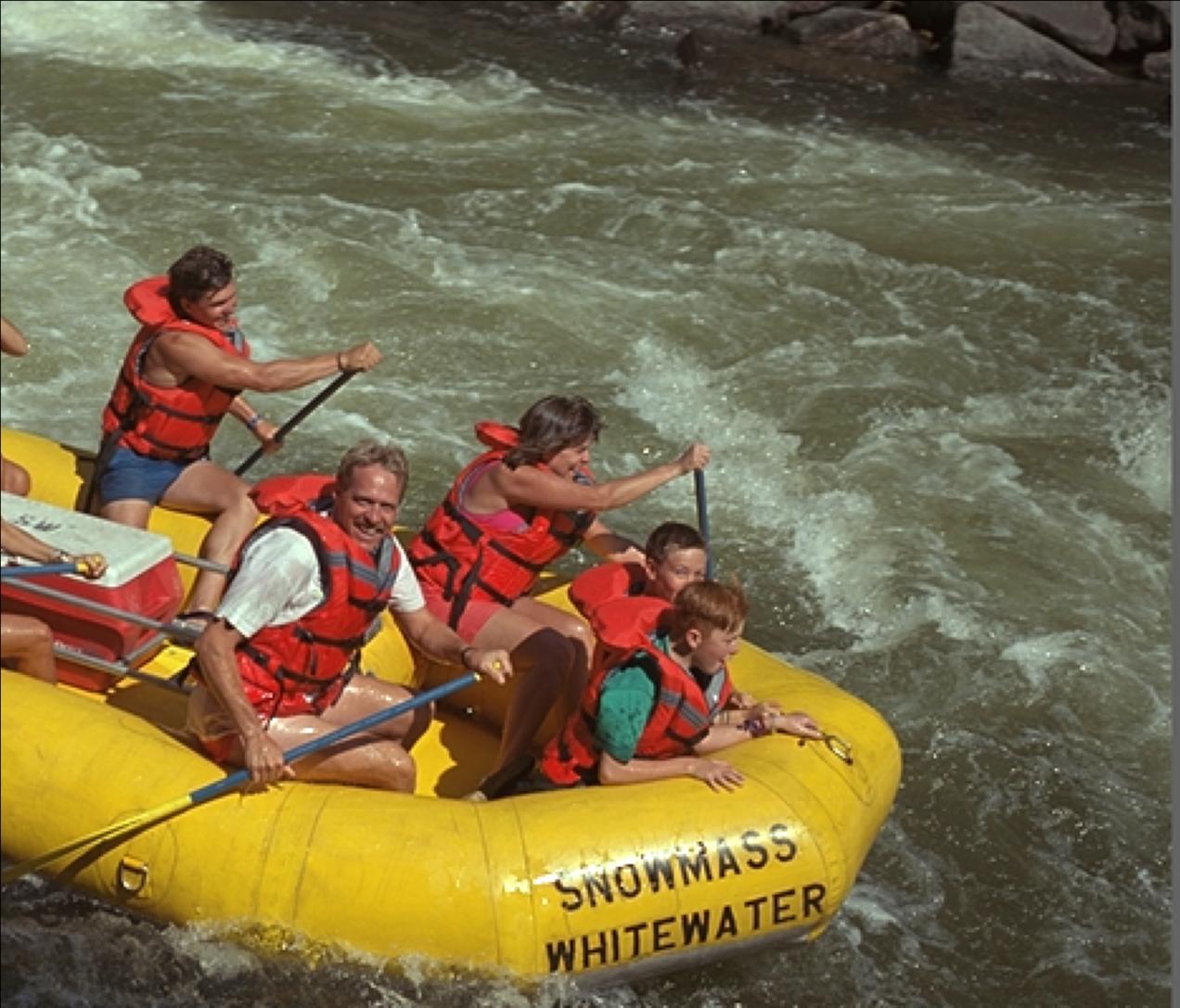}
    \end{minipage}
    \hspace{0.2cm}
    \begin{minipage}[b]{0.21\linewidth}
    \centering
    \centerline{\small Base+Enh. (28.8dB / 0.467bpp)}  
    \vspace{0.1cm}
    \includegraphics[width=\textwidth]{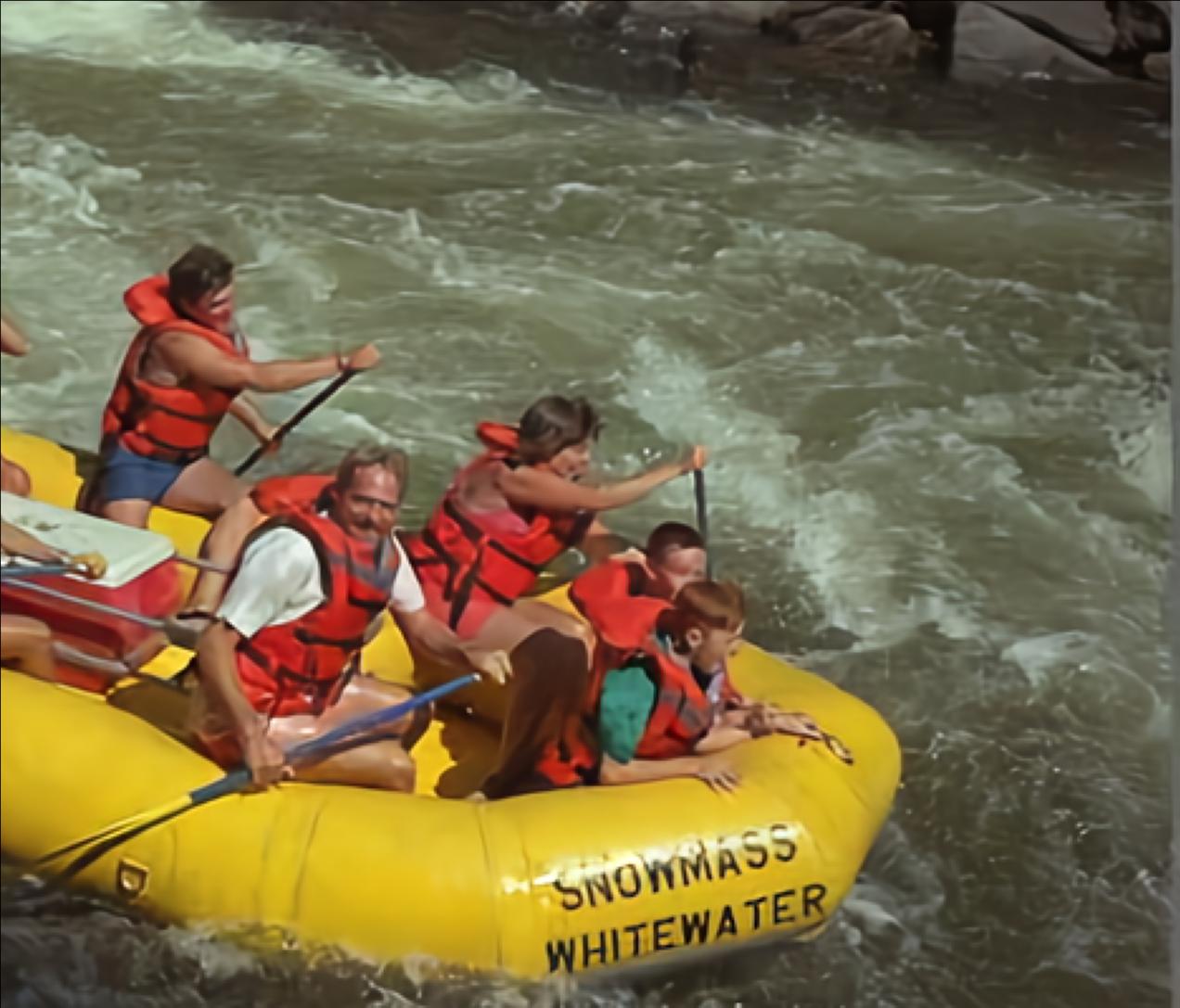}
    \end{minipage}
    \hspace{0.2cm}
    \begin{minipage}[b]{0.21\linewidth}
    \centering
    \centerline{\small Base-only (23.6dB / 0.184bpp)}  
    \vspace{0.1cm}
    \includegraphics[width=\textwidth]{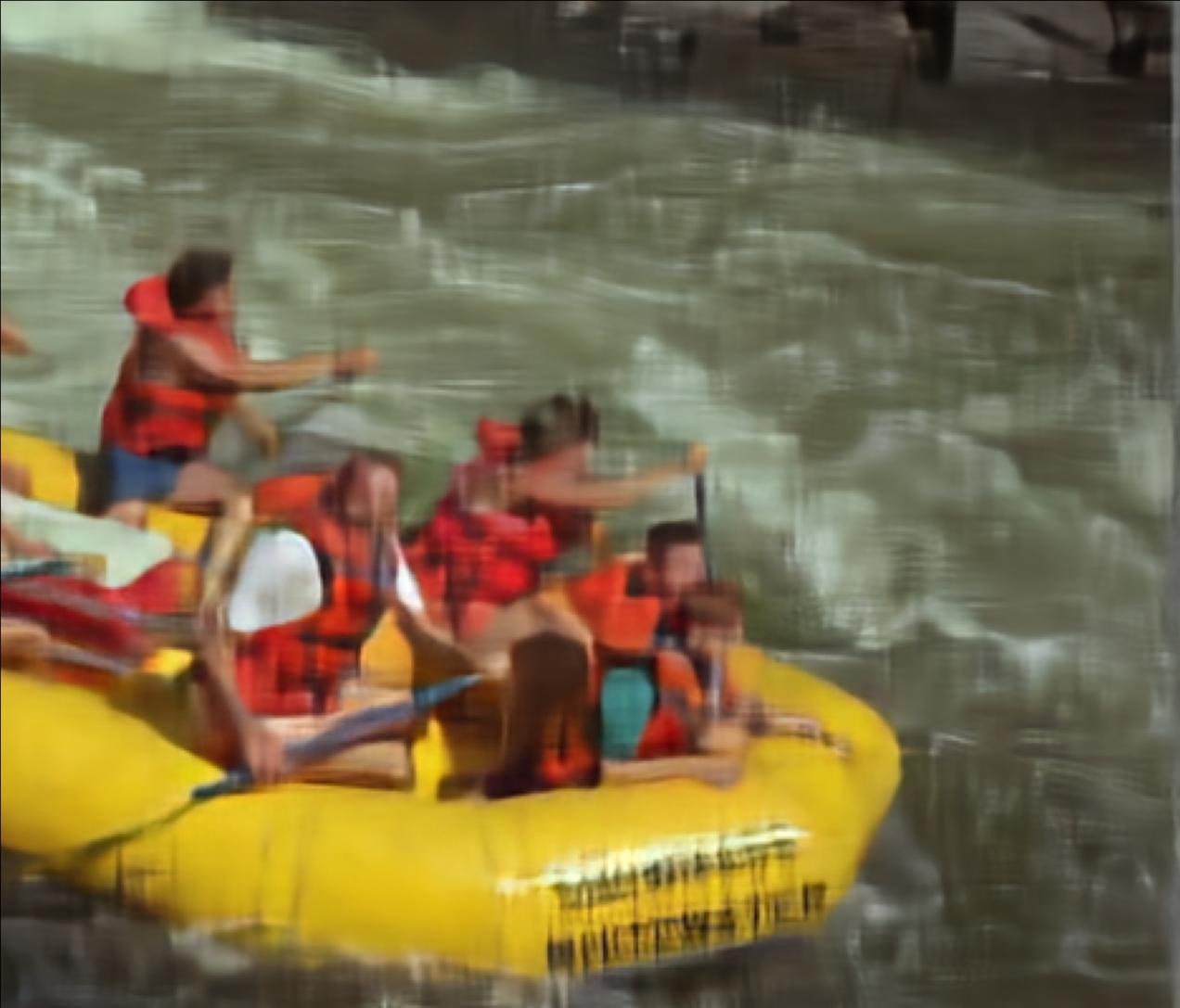}
    \end{minipage}
    \hspace{0.2cm}
    \begin{minipage}[b]{0.21\linewidth}
    \centering
    \centerline{\small Base+ROIs (25.2dB / 0.288bpp)}  
    \vspace{0.1cm}
    \includegraphics[width=\textwidth]{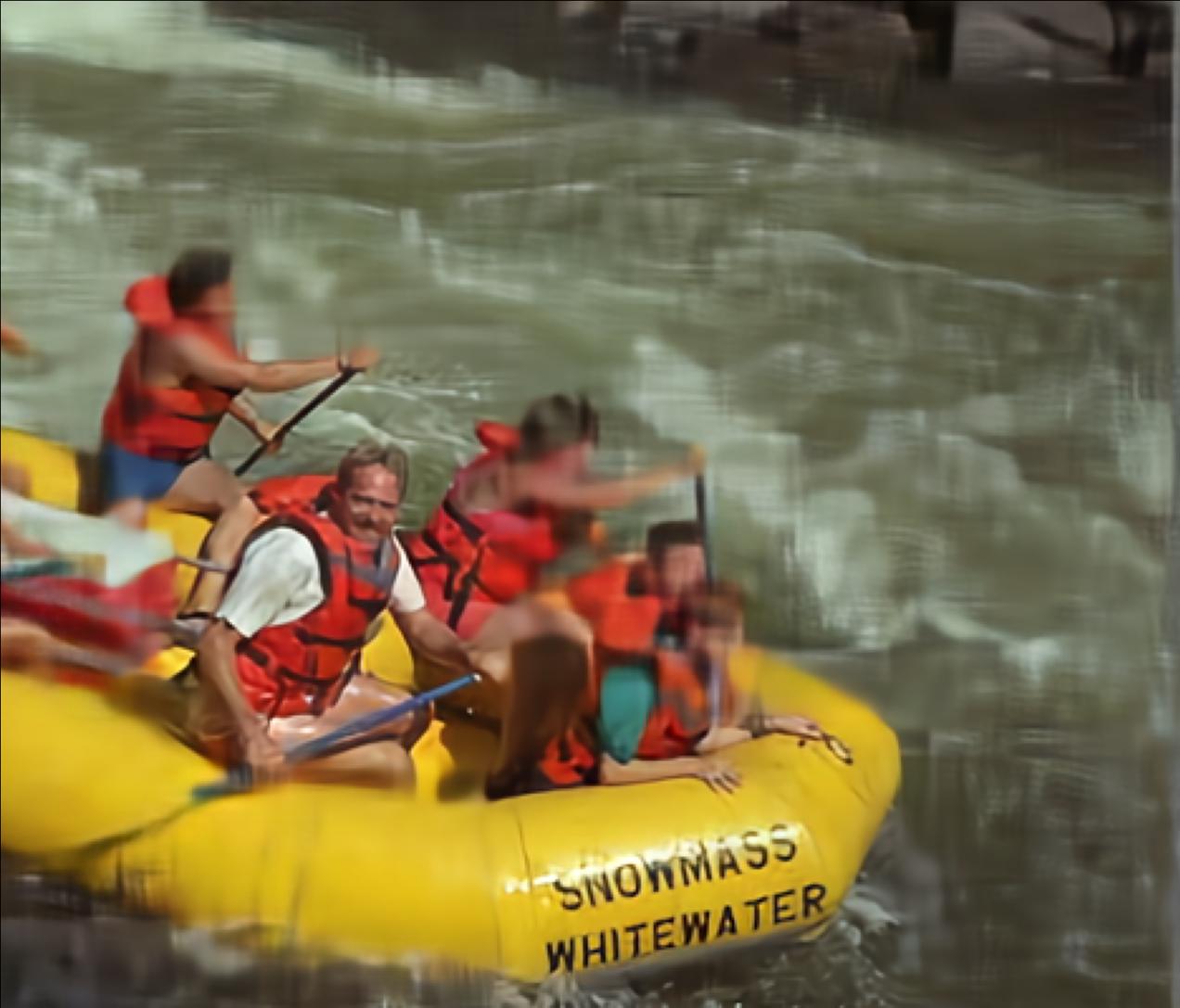}
    \end{minipage}

    \vspace{0.1cm}
    \centering
    \begin{minipage}[b]{0.21\linewidth}
    \centering
    \includegraphics[width=\textwidth]{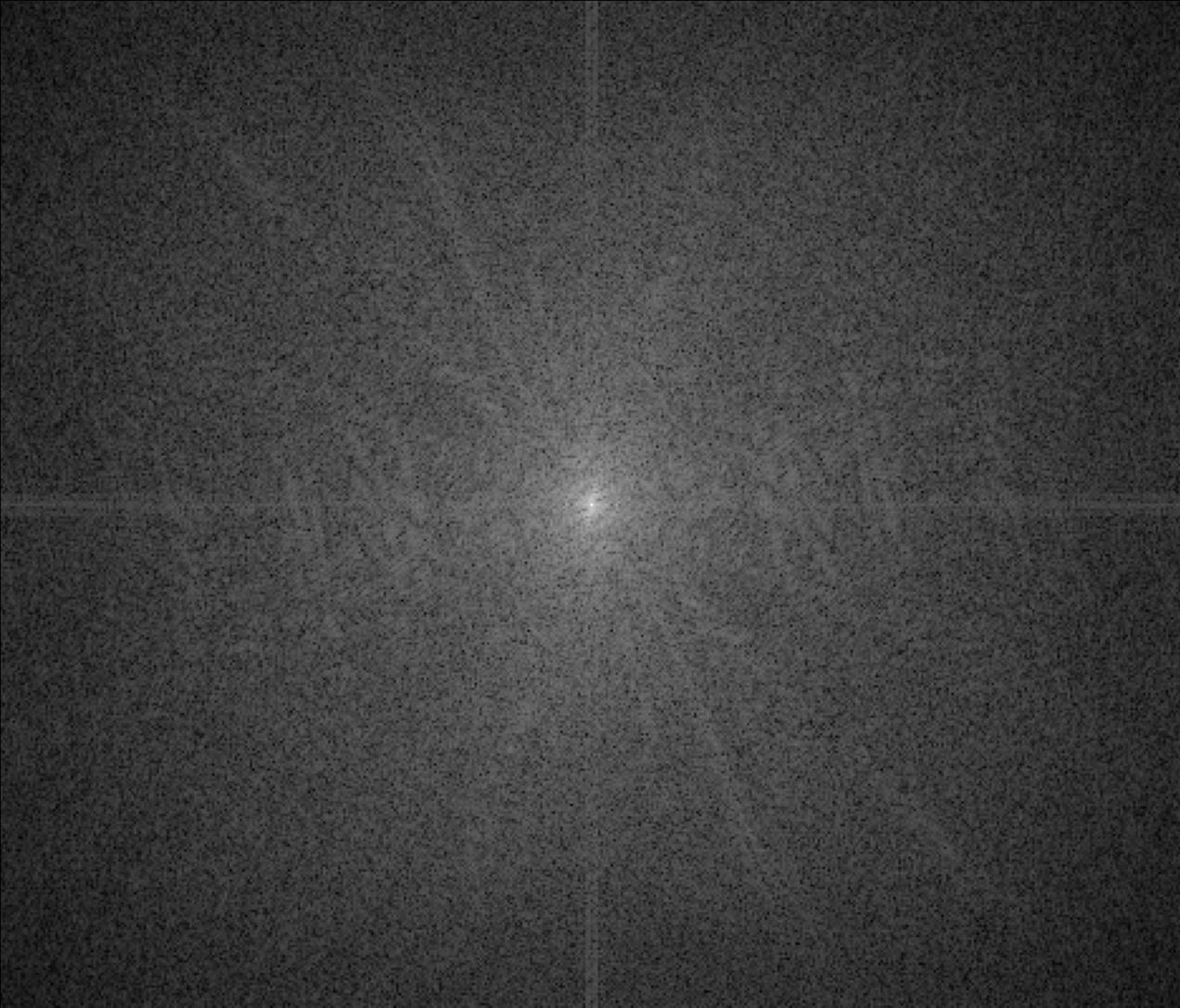}
    \end{minipage}
    \hspace{0.2cm}
    \begin{minipage}[b]{0.21\linewidth}
    \centering
    \includegraphics[width=\textwidth]{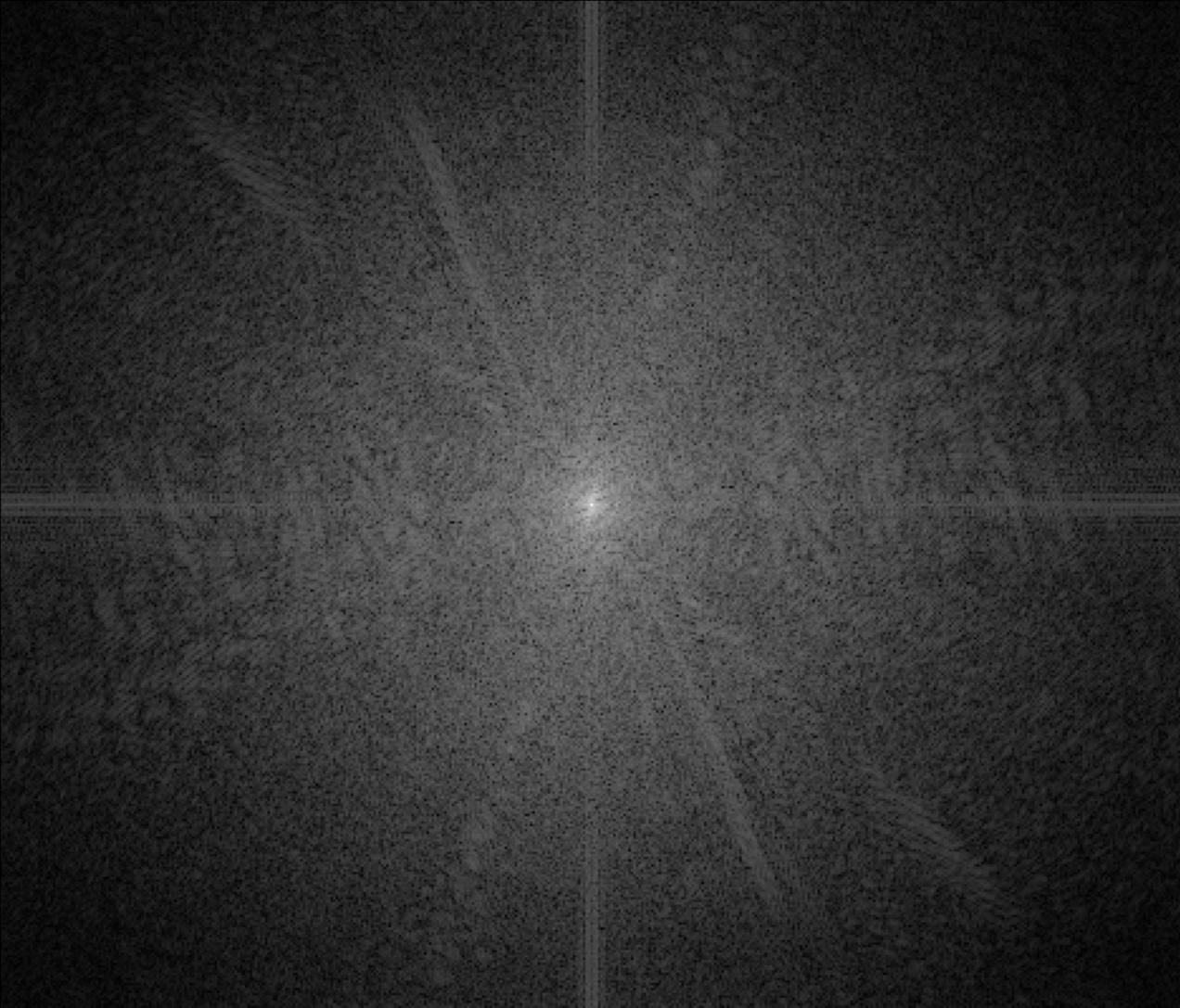}
    \end{minipage}
    \hspace{0.2cm}
    \begin{minipage}[b]{0.21\linewidth}
    \centering
    \includegraphics[width=\textwidth]{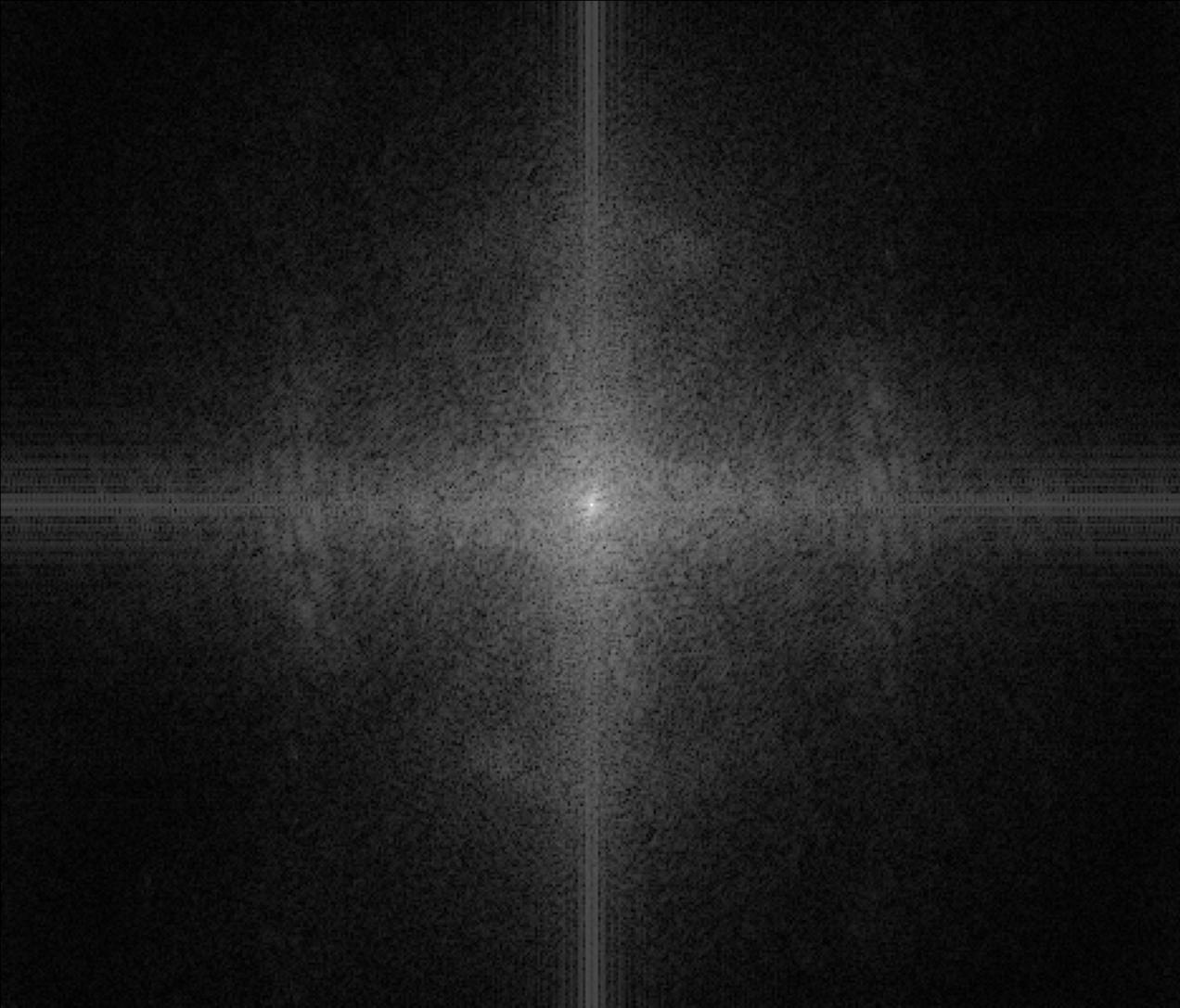}
    \end{minipage}
    \hspace{0.2cm}
    \begin{minipage}[b]{0.21\linewidth}
    \centering
    \includegraphics[width=\textwidth]{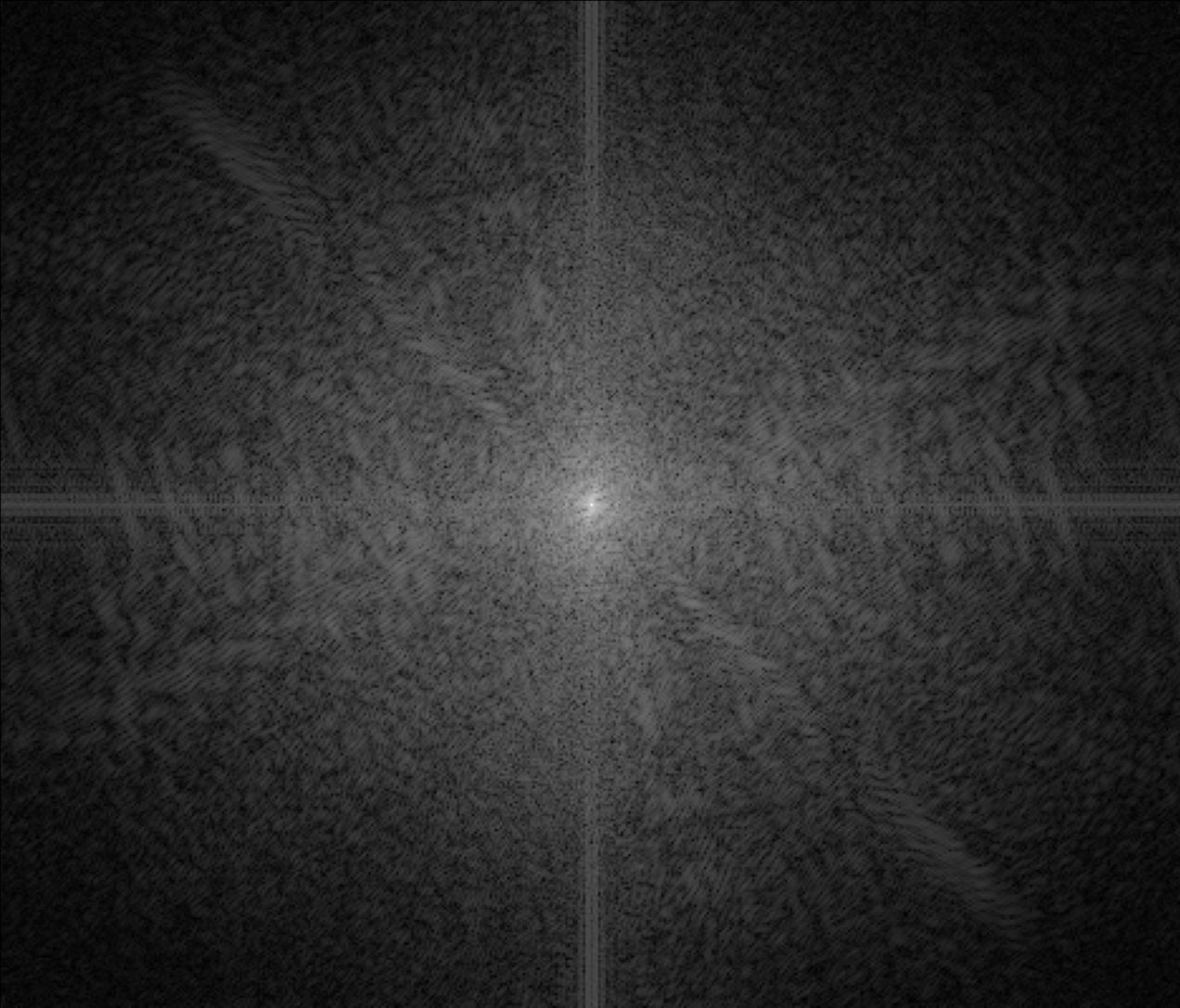}
    \end{minipage}
    
    \vspace{0.1cm}
    \centering
    \begin{minipage}[b]{0.068\linewidth}
    \centering
    \includegraphics[width=\textwidth]{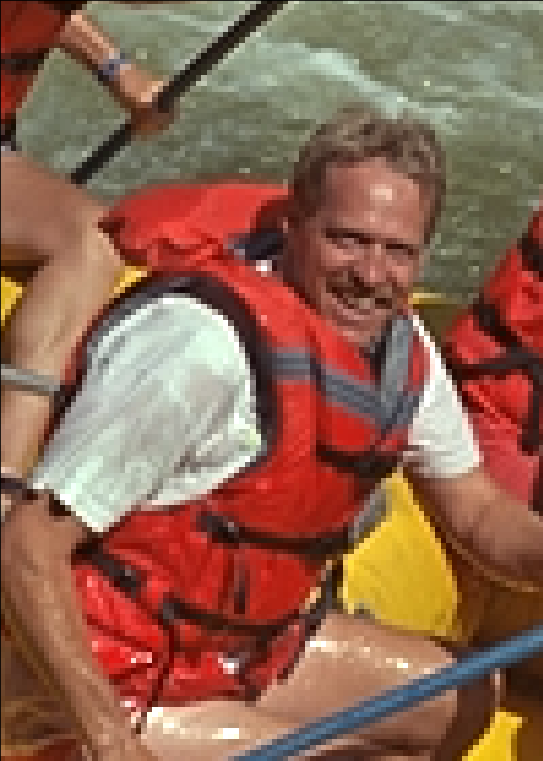}
    \end{minipage}
    \begin{minipage}[b]{0.134\linewidth}
    \centering
    \includegraphics[width=\textwidth]{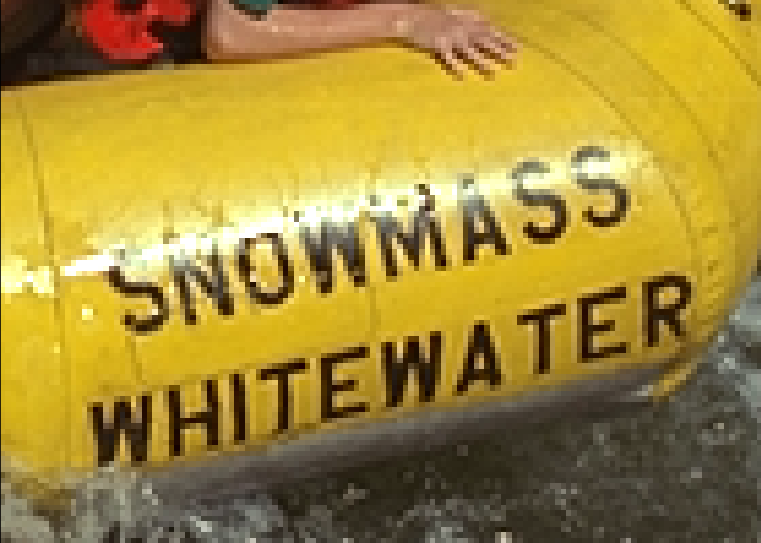}
    \end{minipage}
    \hspace{0.232cm}
    \begin{minipage}[b]{0.068\linewidth}
    \centering
    \includegraphics[width=\textwidth]{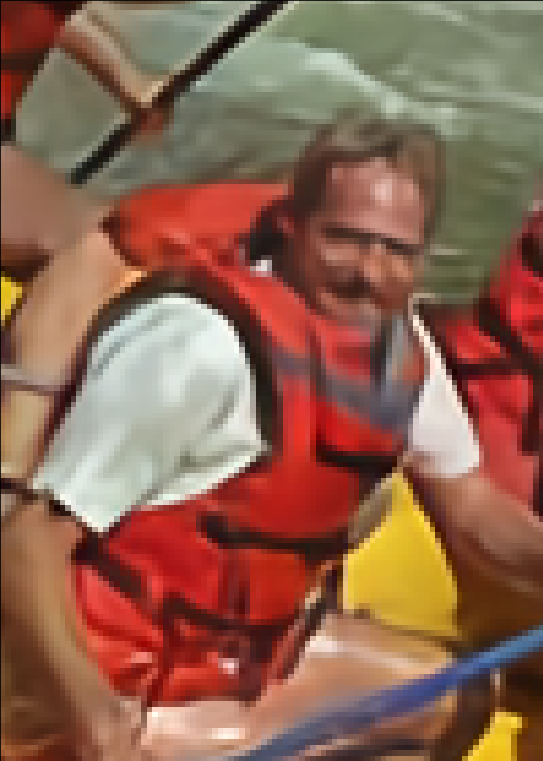}
    \end{minipage}
    \begin{minipage}[b]{0.134\linewidth}
    \centering
    \includegraphics[width=\textwidth]{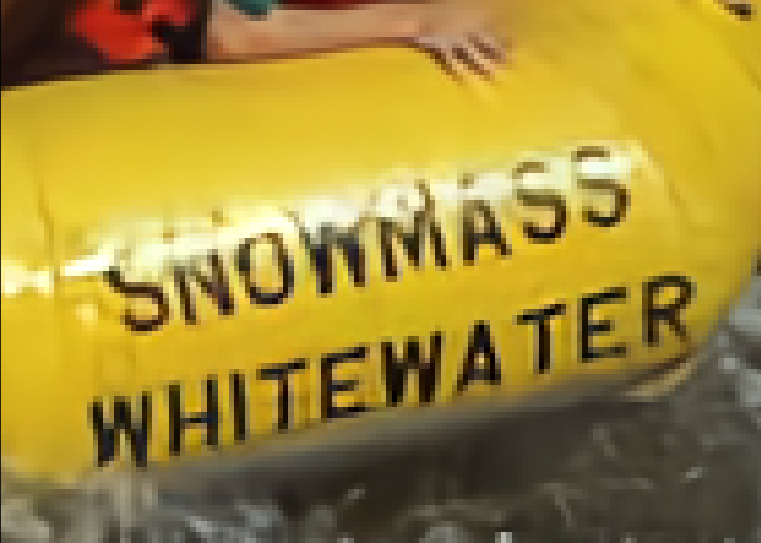}
    \end{minipage}
    \hspace{0.232cm}
    \begin{minipage}[b]{0.068\linewidth}
    \centering
    \includegraphics[width=\textwidth]{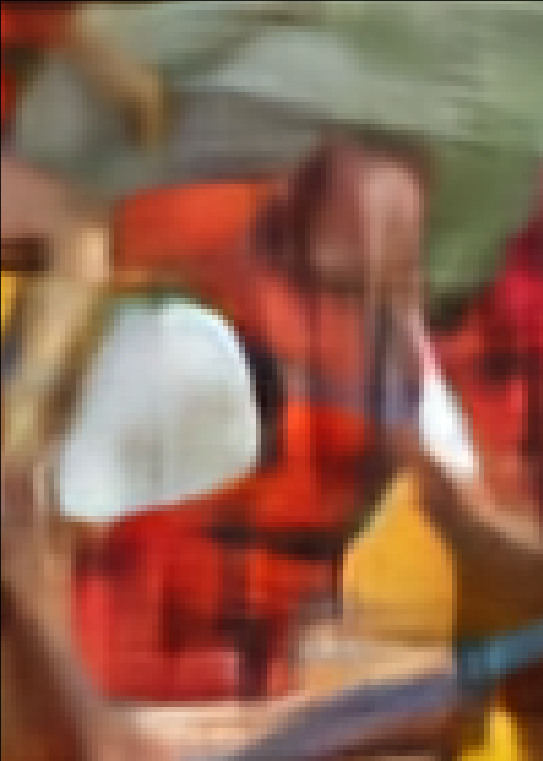}
    \end{minipage}
    \begin{minipage}[b]{0.134\linewidth}
    \centering
    \includegraphics[width=\textwidth]{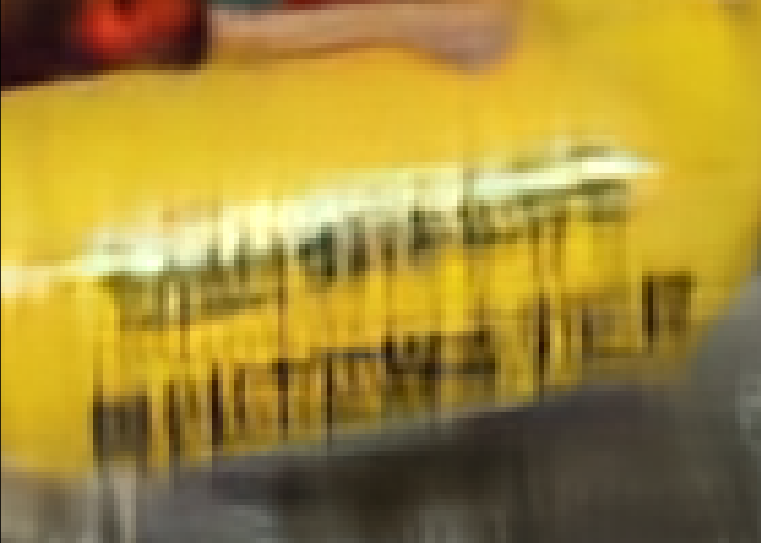}
    \end{minipage}
    \hspace{0.232cm}
    \begin{minipage}[b]{0.068\linewidth}
    \centering
    \includegraphics[width=\textwidth]{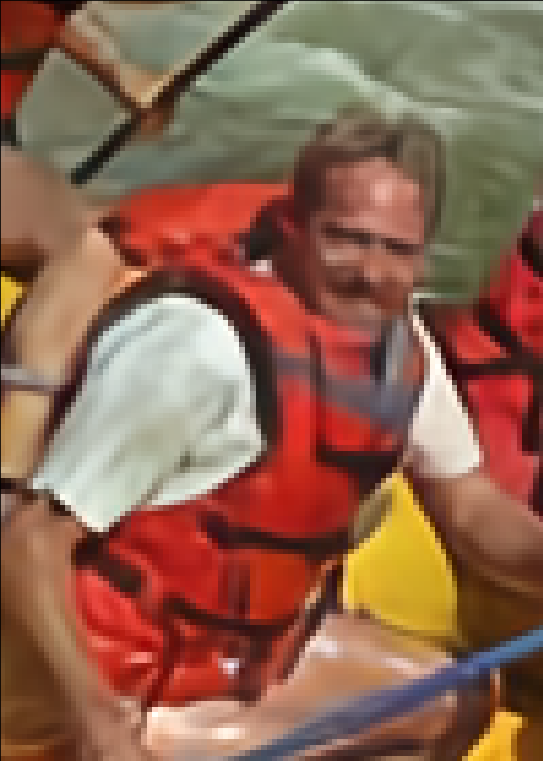}
    \end{minipage}
    \begin{minipage}[b]{0.134\linewidth}
    \centering
    \includegraphics[width=\textwidth]{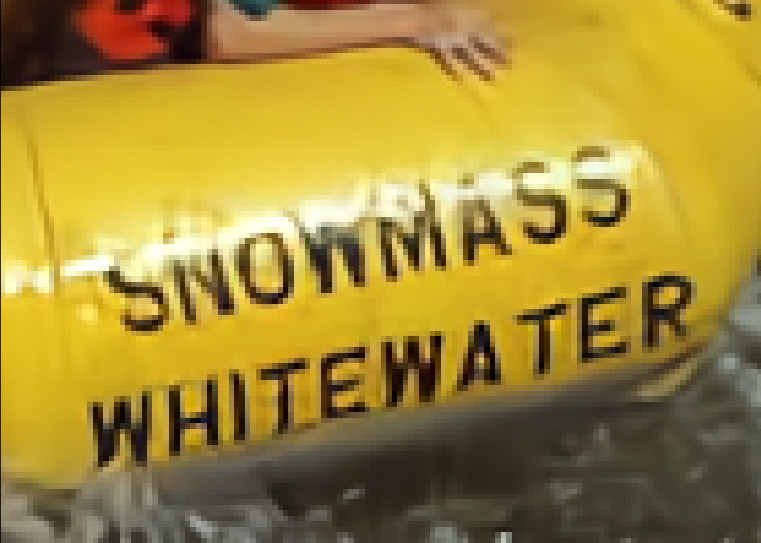}
    \end{minipage}
    
    \vspace{0.1cm}
    \centering
    \begin{minipage}[b]{0.21\linewidth}
    \centering
    \centerline{(a)}\medskip
    \end{minipage}
    \hspace{0.2cm}
    \begin{minipage}[b]{0.21\linewidth}
    \centering
    \centerline{(b)}\medskip
    \end{minipage}
    \hspace{0.2cm}
    \begin{minipage}[b]{0.21\linewidth}
    \centering
    \centerline{(c)}\medskip
    \end{minipage}
    \hspace{0.2cm}
    \begin{minipage}[b]{0.21\linewidth}
    \centering
    \centerline{(d)}\medskip
    \end{minipage}
\vspace{-0.4cm}
\caption{Visual example of our quality scalability method for the sample input image \texttt{kodim14.png} from the Kodak~\cite{kodak_dataset} dataset. The first column (a) represents the uncompressed input. The top header shows the coding results (RGB-PSNR / bpp) for the last three columns in the case of: (b) using both base and enhancement bitstreams, (c) using only the base bitstream, and (d) using the base bitstream along with the enhancement bitstream containing only selected ROIs. For visibility, the first row shows a cropped patch of the compressed image at (300, 50) with a size of $468\times400$. The second row shows the image represented in the Fourier domain. The third row shows enlarged ROIs.}
\label{fig:quality_scalable}
\end{figure*}

\subsubsection{Comparison with benchmarks}
Table~\ref{tbl:bd_performance} summarizes the average bit savings by several LIC methods and our proposed models with various $\alpha$ against JPEG2000%
\footnote{FFMPEG~\cite{tomar2006converting} (v3.4.8) and libopenjpeg~\cite{openjpeg} (v2.3.0) are used.}
in terms of BD-rate~\cite{bd_br}.
Since our method is built upon~\cite{balle2016end} and uses 30M parameters, we reconfigured and retrained the relevant benchmarks~\cite{balle2016end, akbari2021learned} to use the same number of parameters and entropy bottleneck as ours.
As such, we reasonably compare the coding results of the benchmarks in the first two columns with our proposed methods in the last three columns of Table~\ref{tbl:bd_performance}.
The first row shows the average bit savings of the models optimized for MSE in terms of BD-rate computed on the PSNR versus bpp curves. The last row shows coding gain by the models optimized for MS-SSIM in terms of BD-rate computed on the MS-SSIM versus bpp curves.


Conventional scalable codecs cost 15-25\% overhead bits to add a scalability layer in comparison with equivalent non-scalable codecs~\cite{shvc_overview}.
Compared to~\cite{balle2016end}, which optimized for MSE, our two-level quality-scalable model with $\alpha=0.1$ only increases bitrate usage by 1.3\%.
Furthermore, for $\alpha=0.1$, our codec outperforms our reimplementation of the non-scalable OctConv model in~\cite{akbari2021learned}.
For the second row comparing models optimized for MS-SSIM,
the OctConv-based LIC models show coding gains compared to~\cite{balle2016end}, which uses only regular convolution.
We attribute these gains to the structural ability of OctConv in efficiently capturing the spatial structure of their input in a manner reminiscent of Wavelet transforms.


\subsection{Quality scalability and frequency analysis}
Fig.~\ref{fig:quality_scalable} shows a visual example of quality scalablility in action, where
an input image $\mathbf{X}$ is compressed using our MSE-optimized FLIC model trained with small $\lambda=0.01$ and $\alpha=0.01$ to make the quality degradation more apparent.
Each of the columns shows an image, its Fourier domain, and two select enlarged ROIs.
The image in each column in Fig.~\ref{fig:quality_scalable} is:
(a)~$\mathbf{X}$, the original uncompressed image,
(b)~$\widehat{\mathbf{X}}$, reconstructed using both base and enhancement bitstreams,
(c)~$\widehat{\mathbf{X}}_{\textup{base}}$, reconstructed using only the base bitstream, and
(d)~$\widehat{\mathbf{X}}_{\textup{ROI}}$, reconstructed using the base bitstream along with the enhancement bitstream containing only selected ROIs.
Since the base bitstream primarily contains low-frequency information, much of the high-frequency information is missing from $\widehat{\mathbf{X}}_{\textup{base}}$. 
Indeed, the Fourier domain for $\widehat{\mathbf{X}}_{\textup{base}}$ shows much less energy in the high-frequency spectrum compared to $\mathbf{X}$ and $\widehat{\mathbf{X}}$.

To demonstrate our model's capability for ROI-based quality scalability, we select two ROIs for enhancement: the man in the center of the boat, and the text on the boat.
These regions are blocky and unreadable in $\widehat{\mathbf{X}}_{\textup{base}}$.
By including these regions within the enhancement bitstream, the decoder effectively reconstructs these regions within $\widehat{\mathbf{X}}_{\textup{ROI}}$, with a quality equivalent to the same regions in $\widehat{\mathbf{X}}$ by PSNR.

\section{Conclusion}
\label{sec:conclusion}
We presented a novel frequency-aware learned image compression (FLIC) framework that uses our newly introduced WeOctConv layer.
The WeOctConv layer is designed to optimize the separation of spatial frequencies into two latent representations, enabling our FLIC to serve two-level quality-scalable coding with minimal overhead bits.
Furthermore, our method efficiently balances the amount of information between the low- and high-frequency latent representations using a scale factor during training.
This allows it to achieve flexible RD trade-offs for the base bitstream while having minimal impact on overall coding gain.
Finally, we demonstrated the potential of our approach in the context of ROI-based quality enhancement by utilizing partial information from the enhancement latent representation, without requiring any extra retraining.

\bibliographystyle{IEEEbib-abbrev} 

\bibliography{refs}

\begin{thebibliography}{10}

\bibitem{sayood2017introduction}
K. Sayood,
\newblock {\em Introduction to data compression},
\newblock Morgan Kaufmann, 2017.

\bibitem{shapiro1993embedded}
J.~M. Shapiro,
\newblock ``Embedded image coding using zerotrees of wavelet coefficients,''
\newblock {\em IEEE Trans. Signal Process.}, vol. 41, no. 12, pp. 3445--3462,
  1993.

\bibitem{christopoulos2000jpeg2000}
C. Christopoulos, A. Skodras, and T. Ebrahimi,
\newblock ``The jpeg2000 still image coding system: an overview,''
\newblock {\em IEEE Trans. Consum. Electron.}, vol. 46, no. 4, pp. 1103--1127,
  2000.

\bibitem{schwarz2007overview}
H. Schwarz, D. Marpe, and T. Wiegand,
\newblock ``Overview of the scalable video coding extension of the h. 264/avc
  standard,''
\newblock {\em IEEE Trans. Circuits Syst. Video Technol.}, vol. 17, no. 9, pp.
  1103--1120, 2007.

\bibitem{boyce2015overview}
J.~M. Boyce, Y. Ye, J. Chen, and A.~K. Ramasubramonian,
\newblock ``Overview of shvc: Scalable extensions of the high efficiency video
  coding standard,''
\newblock {\em IEEE Trans. Circuits Syst. Video Technol.}, vol. 26, no. 1, pp.
  20--34, 2015.

\bibitem{balle2015density}
J. Ball{\'e}, V. Laparra, and E.~P. Simoncelli,
\newblock ``Density modeling of images using a generalized normalization
  transformation,''
\newblock in {\em Proc. ICLR}, 2016.

\bibitem{Goodfellow-et-al-2016}
G. Ian, B. Yoshua, and C. Aaron,
\newblock {\em Deep Learning},
\newblock MIT Press, 2016.

\bibitem{balle2016end}
J. Ball{\'e}, V. Laparra, and E.~P. Simoncelli,
\newblock ``End-to-end optimized image compression,''
\newblock in {\em Proc. ICLR}, 2017.

\bibitem{balle2018variational}
J. Ball{\'e}, D. Minnen, S. Singh, S.~J. Hwang, and N. Johnston,
\newblock ``Variational image compression with a scale hyperprior,''
\newblock in {\em Proc. ICLR}, 2018.

\bibitem{minnen2018joint}
D. Minnen, J. Ball{\'e}, and G.~D. Toderici,
\newblock ``Joint autoregressive and hierarchical priors for learned image
  compression,''
\newblock {\em Adv. Neural Inf. Process. Syst.}, vol. 31, 2018.

\bibitem{cheng2020image}
Z. Cheng, H. Sun, M. Takeuchi, and J. Katto,
\newblock ``Learned image compression with discretized gaussian mixture
  likelihoods and attention modules,''
\newblock in {\em Proc. IEEE CVPR}, 2020.

\bibitem{hevc_std}
{Int. Telecommun. Union-Telecommun. (ITU-T) and Int. Standards
  Org./Int/Electrotech. Commun. (ISO/IEC JTC 1)},
\newblock ``High efficiency video coding,'' Rec. ITU-T H.265 and ISO/IEC
  23008-2, 2019.

\bibitem{vvc_std}
{Int. Telecommun. Union-Telecommun. (ITU-T) and Int. Standards
  Org./Int/Electrotech. Commun. (ISO/IEC JTC 1)},
\newblock ``Versatile video coding,'' Rec. ITU-T H.266 and ISO/IEC 23090-3,
  2020.

\bibitem{xie2021enhanced}
Y. Xie, K.~L. Cheng, and Q. Chen,
\newblock ``Enhanced invertible encoding for learned image compression,''
\newblock in {\em Proc. ACM Int. Conf. Multimed.}, 2021.

\bibitem{akbari2021learned}
M. Akbari, J. Liang, J. Han, and C. Tu,
\newblock ``Learned bi-resolution image coding using generalized octave
  convolutions,''
\newblock in {\em Proc. AAAI}, 2021, pp. 6592--6599.

\bibitem{chen2019drop}
Y. Chen, H. Fan, B. Xu, Z. Yan, Y. Kalantidis, M. Rohrbach, S. Yan, and J.
  Feng,
\newblock ``Drop an octave: Reducing spatial redundancy in convolutional neural
  networks with octave convolution,''
\newblock in {\em Proc. IEEE ICCV}, 2019, pp. 3435--3444.

\bibitem{wolter2021adaptive}
M. Wolter and J. Garcke,
\newblock ``Adaptive wavelet pooling for convolutional neural networks,''
\newblock in {\em Int. Conf. Artif. Intell. Stat.} PMLR, 2021, pp. 1936--1944.

\bibitem{shi2016real}
W. Shi, J. Caballero, F. Husz{\'a}r, J. Totz, A.~P. Aitken, R. Bishop, D.
  Rueckert, and Z. Wang,
\newblock ``Real-time single image and video super-resolution using an
  efficient sub-pixel convolutional neural network,''
\newblock in {\em Proc. IEEE CVPR}, 2016, pp. 1874--1883.

\bibitem{begaint2020compressai}
J. B{\'e}gaint, F. Racap{\'e}, S. Feltman, and A. Pushparaja,
\newblock ``Compressai: a pytorch library and evaluation platform for
  end-to-end compression research,''
\newblock {\em arXiv preprint arXiv:2011.03029}, 2020.

\bibitem{kodak_dataset}
E. Kodak,
\newblock ``Kodak lossless true color image suite ({P}hoto{CD} {PCD}0992),''
  \url{http://r0k.us/graphics/kodak}.

\bibitem{wang2003multiscale}
Z. Wang, E.~P. Simoncelli, and A.~C. Bovik,
\newblock ``Multiscale structural similarity for image quality assessment,''
\newblock in {\em Proc. IEEE Asilomar Conf. Signals, Systems \& Computers},
  2003, vol.~2, pp. 1398--1402.

\bibitem{xue2019video}
T. Xue, B. Chen, J. Wu, D. Wei, and W.~T. Freeman,
\newblock ``Video enhancement with task-oriented flow,''
\newblock {\em Int. J. Comput. Vision}, vol. 127, no. 8, pp. 1106--1125, 2019.

\bibitem{tomar2006converting}
S. Tomar,
\newblock ``Converting video formats with ffmpeg,''
\newblock {\em Linux Journal}, vol. 2006, no. 146, pp. 10, 2006.

\bibitem{openjpeg}
``{JPEG}2000 reference software,'' [Online]:
  \url{https://github.com/uclouvain/openjpeg},
\newblock Accessed: 2022-05-04.

\bibitem{bd_br}
G. Bj{\o}ntegaard,
\newblock ``{VCEG-M33}: Calculation of average {PSNR} differences between {RD}
  curves,''
\newblock in {\em Video Coding Experts Group (VCEG)}, Apr. 2001.

\bibitem{shvc_overview}
J.~M. {Boyce}, Y. {Ye}, J. {Chen}, and A.~K. {Ramasubramonian},
\newblock ``Overview of {SHVC:} scalable extensions of the high efficiency
  video coding standard,''
\newblock {\em IEEE Trans. Circuits Syst. Video Technol.}, vol. 26, no. 1, pp.
  20--34, 2016.

\end{thebibliography}

\end{document}